\RequirePackage{fix-cm}
\documentclass[twocolumn,epjc3]{svjour3}  

\usepackage{graphicx}
\usepackage{epstopdf}
\usepackage{multirow}
\usepackage{soul}
\usepackage{dcolumn}
\usepackage{bm}
\usepackage{mathrsfs}
\usepackage[colorlinks,citecolor=blue,anchorcolor=red,menucolor=red,linkcolor=red,filecolor=red,runcolor=red,urlcolor=blue,frenchlinks=red]{hyperref}
\usepackage{color}
\usepackage{array}
\usepackage{longtable}
\usepackage{subfigure}
\usepackage{float}

\newcommand{\mev}{\textrm{ MeV}}

\smartqed  
\RequirePackage{graphicx}
\journalname{Eur. Phys. J. C}
\begin{document}

\title{Fully-Heavy Hadronic Molecules $B_c^{(*)+} B_c^{(*)-}$ Bound by Fully-Heavy Mesons
}


\author{Wen-Ying Liu\thanksref{addr1}
        \and
        Hua-Xing Chen\thanksref{e1,addr1} 
}

\thankstext{e1}{e-mail: hxchen@seu.edu.cn}


\institute{School of Physics, Southeast University, Nanjing 210094, China \label{addr1}
}

\date{Received: date / Accepted: date}

\maketitle

\begin{abstract}
A lot of exotic hadrons have been reported in the past twenty years, which bring us the renaissance of the hadron spectroscopy. Most of them can be understood as hadronic molecules, whose interactions are mainly due to the exchange of light mesons, and specifically, light vector mesons through the coupled-channel unitary approach within the local hidden-gauge formalism. It is still controversial whether the interaction arising from the exchange of heavy mesons is capable of forming hadronic molecules. We apply the coupled-channel unitary approach to study the fully-heavy $b \bar b c \bar c$ system, where the exchanged mesons can only be the fully-heavy vector mesons $J/\psi$, $B_c^*$, and $\Upsilon$. Especially, the $J/\psi$ meson is much lighter than the $B_c^{(*)}$ mesons, so the present study can be taken as a general investigation on the question whether a lower-mass fully-heavy meson is able to bind two higher-mass fully-heavy hadrons. Our results suggest the existence of the fully-heavy hadronic molecules $|B_c^{+} B_c^{-}; J^{PC}=0^{++} \rangle$, $|B_c^{*+} B_c^{-} - c.c.; J^{PC}=1^{+-} \rangle$, and $|B_c^{*+} B_c^{*-}; J^{PC}=2^{++} \rangle$ as well as the possible existence of $|B_c^{*+} B_c^{-} + c.c.; J^{PC}=1^{++} \rangle$. These states are potential to be observed in the $J/\psi \Upsilon$, $\mu^+ \mu^- J/\psi$, and $\mu^+ \mu^- \Upsilon$ channels in future ATLAS, CMS, and LHCb experiments.
\end{abstract}

\section{Introduction}
\label{sec:intro}
The exotic hadrons, such as compact multiquarks and hadronic molecules, can not be explained in the conventional quark model as $q \bar q$ mesons and $qqq$ baryons. Studies on these states have received much attention in the past twenty years and become a crucial subject in hadron physics. Since 2015 many hidden-charm pentaquark states have been discovered by the LHCb Collaboration, including the $P^N_\psi(4312)^+$, $P^N_\psi(4440)^+$, $P^N_\psi(4457)^+$, $P^\Lambda_{\psi s}(4338)^0$, and $P^\Lambda_{\psi s}(4459)^0$~\cite{LHCb:2015yax,LHCb:2019kea,LHCb:2020jpq,LHCb:2022ogu}. These structures are just below the $\bar D^{(*)} \Sigma_c$ and $\bar D^{(*)} \Xi_c$ thresholds, so it is natural to explain them as the $\bar D^{(*)} \Sigma_c$ and $\bar D^{(*)} \Xi_c$ hadronic molecules, whose existence has been predicted in Refs.~\cite{Wu:2010jy,Chen:2015sxa,Xiao:2019gjd,Wang:2011rga,Yang:2011wz,Karliner:2015ina,Cheng:2015cca,Santopinto:2016pkp,Chen:2016ryt,Shen:2019evi,Wang:2019nvm,Shen:2020gpw} through various theoretical methods.

The coupled-channel unitary approach was initially developed to study dynamically generated low-lying scalar hadrons~\cite{Oller:1997ti,Oller:1998zr,GomezNicola:2001as} and baryons~\cite{Oller:2000fj,Jido:2003cb}, and was subsequently extended to higher energy regions~\cite{Hofmann:2003je,Geng:2008gx,Guo:2006fu,Mizutani:2006vq,Lu:2016gev}. Especially, the coupled-channel unitary approach within the local hidden-gauge formalism was extensively applied in Refs.~\cite{Wu:2010jy,Chen:2015sxa,Xiao:2019gjd} to study the hidden-charm pentaquark states. Besides, this method has been widely and successfully applied to study many other hadronic molecules that contain some light quarks, where the interactions mainly arise from the exchange of light vector mesons~\cite{Chen:2022asf,Guo:2017jvc}. Oppositely, in the fully-heavy multiquark system, the exchanged mesons can only be the fully-heavy vector mesons within this approach, and it is controversial whether the induced interaction can be large enough to form the fully-heavy hadronic molecules.

Actually, since 2020 the LHCb, CMS, and ATLAS collaborations have observed several exotic structures in the di-$J/\psi$ invariant mass spectrum~\cite{LHCb:2020bwg,ATLAS:2023bft,CMS:2023owd}, including the $X(6200)$, $X(6600)$, $X(6900)$, and $X(7200)$. These structures are good candidates for the fully-charmed tetraquark states, and especially, the LHCb measurement~\cite{LHCb:2020bwg} is in a remarkable coincidence with the QCD sum rule
results of Ref.~\cite{Chen:2016jxd}, suggesting that the $X(6900)$ can be interpreted as a $P$-wave compact fully-charmed tetraquark state. Some theorists have attempted to explain them as the hadronic molecules composed of two charmonia~\cite{Czarnecki:2017vco,Guo:2020pvt,Cao:2020gul,Gong:2020bmg,Dong:2021lkh}, but there does not exist a promising answer to this question.

In order to search for the possibly-existing fully-heavy hadronic molecules, in this letter we apply the coupled-channel unitary approach within the local hidden gauge formalism to study the $b \bar b c \bar c$ system. This is a fully-heavy system, where the exchanged mesons can only be the fully-heavy vector mesons $J/\psi$, $B_c^{*\pm}$, and $\Upsilon$. Note that there have been some theoretical studies on the compact tetraquark states with the quark content $b \bar b c \bar c$~\cite{Wu:2016vtq,Richard:2017vry,Anwar:2017toa,Yang:2021zrc,Wang:2021mma,Zhang:2022qtp}, but such a system has not been extensively investigated within the hadronic molecular picture yet, so it also provides an excellent opportunity to test the applicability of our approach itself.

In this letter we investigate altogether eight coupled channels, including $\eta_c\eta_b$, $J/\psi\eta_b$, $\Upsilon\eta_c$, $J/\psi\Upsilon$, $B_c^+B_c^-$, $B_c^{*+}B_c^-$, $B_c^{*-}B_c^{+}$, and $B_c^{*+}B_c^{*-}$. After deriving their scattering amplitudes and solving the Bethe-Salpeter equation, we find the possible existence of four poles that can qualify as the fully-heavy hadronic molecules composed of the $B_c^{(*)+}$ and $B_c^{(*)-}$ mesons. Among them, $|B_c^{+} B_c^{-}; J^{PC}=0^{++} \rangle$, $|B_c^{*+} B_c^{-} - c.c.; J^{PC}=1^{+-} \rangle$, and $|B_c^{*+} B_c^{*-}; J^{PC}=2^{++} \rangle$ are more likely to exist, where the coupled-channel effects are important. Besides, $|B_c^{*+} B_c^{-} + c.c.; J^{PC}=1^{++} \rangle$ may exist or it may also behave as a threshold cusp, whose interaction is mainly due to the exchange of the $J/\psi$ meson. Their potential observation channels are $J/\psi \Upsilon$, $\mu^+ \mu^- J/\psi$, and $\mu^+ \mu^- \Upsilon$ at LHC.
\section{Formalism}
\label{sec:form}
\begin{figure}[H]
    \centering
    \includegraphics[width=1\linewidth]{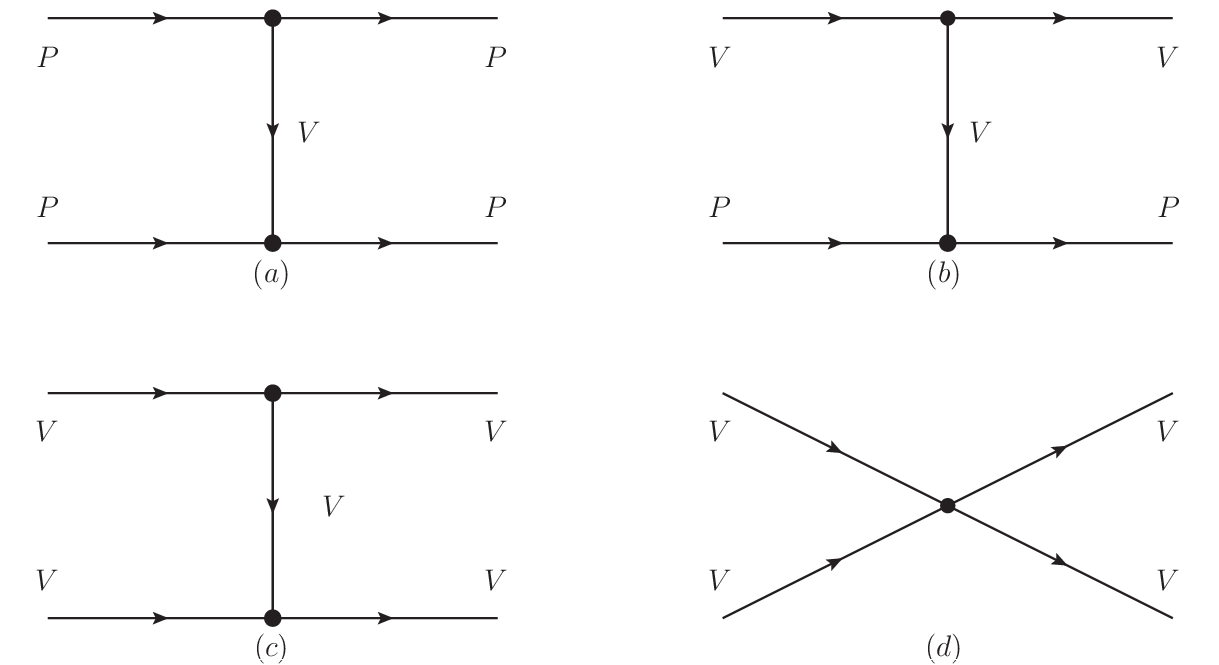}
    \caption{The interactions of the $b \bar b c \bar c$ system arising from: (a) the vector meson exchange between two pseudoscalar mesons, (b) the vector meson exchange between one vector meson and one pseudoscalar meson, (c) the vector meson exchange between two vector mesons, and (d) the contact term connecting four vector mesons.}
    \label{fig-lagrangians}
\end{figure}

Within the local hidden gauge formalism~\cite{Bando:1987br,Meissner:1987ge}, the meson-meson interactions mainly arise from the exchange of vector mesons, as depicted in Fig.~\ref{fig-lagrangians}(a,b,c). Besides, we also take into account the contact term depicted in Fig.~\ref{fig-lagrangians}(d), and their corresponding Lagrangians can be altogether written as:
\begin{eqnarray}
    \nonumber
    \mathcal{L}_{VPP} &=& -ig \, \langle[P,\partial_{\mu} P] V^{\mu}\rangle \, ,
    \\[1.5mm]
    \mathcal{L}_{VVV} &=& ig \, \langle(V^{\mu}\partial_{\nu}V_{\mu}-\partial_{\nu}V^{\mu}V_{\mu})V^{\nu}\rangle \, ,
    \label{eq-lagrangians}
    \\
    \nonumber
    \mathcal{L}_{VVVV} &=& \frac{g^2}{2}\langle V_{\mu}V_{\nu}V^{\mu}V^{\nu}-V_{\nu}V_{\mu}V^{\mu}V^{\nu}\rangle \, .
\end{eqnarray}
It should be noted that the applicability of the extended local hidden gauge formalism to heavy-flavor systems remains a subject of debate. In the local hidden gauge framework~\cite{Bando:1987br,Meissner:1987ge}, vector mesons are treated as gauge bosons that mediate interactions. This mechanism has been successful in describing numerous low-energy interactions~\cite{Wu:2010jy,Geng:2008gx,Oset:2010tof,Aceti:2014uea,Wang:2023jeu}, particularly in processes dominated by the exchange of light vector mesons. However, it remains unclear whether this approach is still valid in the heavy-flavor sector. Consequently, the present study, along with Ref.~\cite{Liu:2024pio}, constitutes pioneering research aimed at exploring fully heavy hadronic molecules within this formalism.

Considering that only the charm and bottom quarks are involved in the present study, we can write the pseudoscalar and vector mesons as two $2\times2$ matrices:
\begin{equation}
    \nonumber
    P =
    \left(\begin{array}{cc}
            \eta_c & B_c^+  \\
            B_c^-  & \eta_b
        \end{array} \right) , ~~~
    V =
    \left( \begin{array}{cc}
            J/\psi       & B_c^{*+} \\
            B_c^{*-}     & \Upsilon
        \end{array} \right) \, .
\end{equation}
The coupling constant $g$ can be generally defined as $g={M_V}/({2f_P})$, where $M_V$ represents the mass of the exchanged vector meson and $f_P$ represents the decay constant of its corresponding pseudoscalar meson. However, the charm and bottom quarks do not form a flavor $SU(2)$ symmetry, so the coupling constant $g$ is not an overall parameter. We respectively use
\begin{eqnarray*}
& M_{J/\psi} = 3096.9 \mbox{ MeV~\cite{pdg},   } f_{\eta_c}=387/\sqrt{2} \mbox{ MeV~\cite{Becirevic:2013bsa}}, &
\\
&  M_{B_c^*} = 6331 \mbox{ MeV~\cite{Mathur:2018epb},   } f_{B_c}=427/\sqrt{2} \mbox{ MeV~\cite{McNeile:2012qf}}, &
\\
& M_{\Upsilon} = 9460.4 \mbox{ MeV~\cite{pdg},   } f_{\eta_b}=667/\sqrt{2} \mbox{ MeV~\cite{McNeile:2012qf}}, &
\end{eqnarray*}
for the exchange of the $J/\psi$, $B_c^*$, and $\Upsilon$ mesons. Besides, we take $g^2 = \sqrt{g_{V_1}g_{V_2}g_{V_3}g_{V_4}}$ for the contact term, with $V_{1\cdots4}$ the four connected vector mesons.

The $PP$ interaction between two pseudoscalar mesons involves two coupled channels $\eta_c \eta_b$ and $B_c^+B_c^-$. The $VV$ interaction between two vector mesons also involves two coupled channels $J/\psi\Upsilon$ and $B_c^{*+}B_c^{*-}$. The $VP$ interaction between vector and pseudoscalar mesons involves four coupled channels, which can be naturally separated into the single channel with the positive $C$-parity:
\begin{equation}
    \nonumber
    B_c^{*} \bar B_c^{(C=+)} \equiv \left(B_c^{*+}B_c^- + c.c. \right)/\sqrt{2}  \, ,
\end{equation}
and three coupled channels with the negative $C$-parity:
\begin{equation}
    \nonumber
    J/\psi\eta_b \, , ~ \Upsilon\eta_c \, , ~ B_c^{*} \bar B_c^{(C=-)} \equiv \left(B_c^{*+}B_c^- - c.c. \right)/\sqrt{2}  \, .
\end{equation}
We use the transition potentials $V_{PP/VP/VV}(s)$ to describe these interactions, which can be derived from Eqs.~(\ref{eq-lagrangians}), as detailedly discussed in the next section. Based on the obtained results, we can further derive the scattering amplitudes by solving the Bethe-Salpeter equation:
\begin{equation}
    T_{PP/VP/VV} = \left({\bf 1} - V_{PP/VP/VV} \bullet G\right)^{-1} \bullet V_{PP/VP/VV} .
    \label{eq-bseq}
\end{equation}
Here $G(s)$ is the diagonal loop function, and its expression for the $i$th channel is
\begin{equation}
    G_{ii}(s) = i\int{\frac{d^4q}{(2\pi)^4}\frac{1}{q^2-m_1^2+i\epsilon}\frac{1}{(p-q)^2-m_2^2+i\epsilon}} \, ,
    \label{eq-G}
\end{equation}
where $s = p^2$ with $p$ the total four-momentum, and $m_{1,2}$ are the masses of the two mesons involved in this channel. We use the cutoff method to regularize it as
\begin{equation}
    G_{ii}(s) = \int_0^{\Lambda} \frac{d^3q}{(2\pi)^3} \frac{\omega_1 + \omega_2}{2\omega_1\omega_2} \frac{1}{s-(\omega_1+\omega_2)^2+i\epsilon} \, ,
    \label{eq-cutoff-G}
\end{equation}
where $\Lambda$ is the cutoff momentum, $\omega_1 = \sqrt{m_1^2+\vec{q}^{\,2}}$, and $\omega_2 = \sqrt{m_2^2+\vec{q}^{\,2}}$.

\section{Interaction Kernel $V$}
\label{sec:kernel}
In this section, we provide some additional details on our framework, especially the specific form of the interaction core $V$. In this letter we use the coupled-channel unitary approach within the local hidden gauge formalism to study the possibly-existing $B_c^{(*)+} B_c^{(*)-}$ hadronic molecules  in the fully-heavy $b \bar b c \bar c$ system. We take into account the $\eta_c\eta_b$, $J/\psi\eta_b$, $\Upsilon\eta_c$, $J/\psi\Upsilon$, $B_c^+B_c^-$, $B_c^{*+}B_c^-$, $B_c^{*-}B_c^{+}$, and $B_c^{*+}B_c^{*-}$ channels. Their threshold masses are summarized in Table~\ref{tab:threshold}. In order to derive the transition potentials $V_{PP/VP/VV}(s)$ from Eqs.~(\ref{eq-lagrangians}), we shall separately investigate in the following subsections: a) the $PP$ interaction between two pseudoscalar mesons, b) the $VP$ interaction between vector and pseudoscalar mesons, and c) the $VV$ interaction between two vector mesons.

\begin{table}[htb]
    \renewcommand{\arraystretch}{1.4}
    \centering
    \caption{Threshold masses of the eight channels considered in the present study, in units of MeV.}
    \setlength{\tabcolsep}{2mm}{
        \begin{tabular}{c|cccc}
            \hline\hline
            Channels  & $\eta_c\eta_b$       & $J/\psi\eta_b$     & $\Upsilon\eta_c$  & $J/\psi\Upsilon$
            \\ \hline
            Threshold & 12382.6              & 12495.6            & 12444.3           & 12557.3
            \\ \hline
            Channels  & $B_c^+B_c^-$         & $B_c^{*+}B_c^-$    & $B_c^{*-}B_c^+$   & $B_c^{*+}B_c^{*-}$
            \\ \hline
            Threshold & 12548.9              & 12605.5            & 12605.5           & 12662.0
            \\ \hline\hline
        \end{tabular}}
    \label{tab:threshold}
\end{table}

\subsection{$PP$ interaction}
\label{sec:kernel-1}

\begin{figure}[h]
    \centering
    \includegraphics[width=1\linewidth]{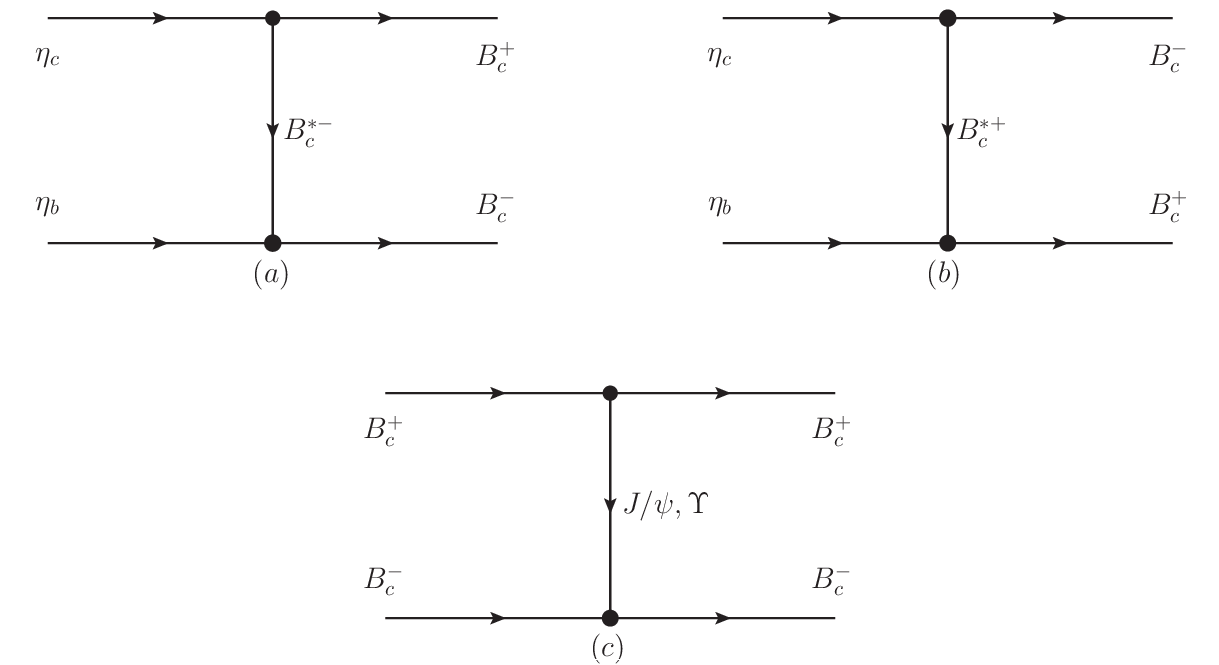}
    \caption{The vector meson exchange between two pseudoscalar mesons: (a) the $t$-channel diagram for $\eta_c\eta_b \to B_c^+B_c^-$, (b) the $u$-channel diagram for $\eta_c\eta_b \to B_c^+B_c^-$, and (c) the $t$-channel diagram for $B_c^+B_c^- \to B_c^+B_c^-$.}
    \label{fig-02}
\end{figure}

In this subsection we investigate the $PP$ interaction arising from the vector meson exchange between two pseudoscalar mesons. There are two coupled channels:
\begin{equation}
    \nonumber
    \eta_c \eta_b \, , ~~~ B_c^+B_c^- \, .
\end{equation}
Their interactions are shown in Fig.~\ref{fig-02}, and the transition potential can be derived from Eqs.~(\ref{eq-lagrangians}) as
\begin{eqnarray}
        V_{PP}(s) &=& C_{PP}^t \times g^2 (p_1+p_3) (p_2+p_4)
        \\ \nonumber &+& C_{PP}^u \times g^2 (p_1+p_4) (p_2+p_3) \, ,
\end{eqnarray}
where $p_1(p_3)$ is the four-momentum of the $\eta_c(B_c^+)$ meson and $p_2(p_4)$ is the four-momentum of the $\eta_b (B_c^-)$ meson.

The two $2\times2$ matrices $C_{PP}^t$ and $C_{PP}^u$ relate to the vector meson exchange in the  $t$- and $u$-channels, respectively:
\begin{eqnarray}
    \setlength{\arraycolsep}{5pt}
    \renewcommand{\arraystretch}{2}
    C_{PP}^t &=& \left(
    \begin{array}{c|cc}
            J=0          & \eta_c\eta_b                 & B_c^+B_c^-
            \\ \hline
            \eta_c\eta_b & 0                            & \lambda\frac{1}{m_{B_c^*}^2}                       \\
            B_c^+B_c^-   & \lambda\frac{1}{m_{B_c^*}^2} & -(\frac{1}{m_{J/\psi}^2}+\frac{1}{m_{\Upsilon}^2})
        \end{array}
    \right)\, ,
\\
    C_{PP}^u &=& \left(
    \begin{array}{c|cc}
            J=0          & \eta_c\eta_b                 & B_c^+B_c^-
            \\ \hline
            \eta_c\eta_b & 0                            & \lambda\frac{1}{m_{B_c^*}^2} \\
            B_c^+B_c^-   & \lambda\frac{1}{m_{B_c^*}^2} & 0
        \end{array}
    \right)\, .
\end{eqnarray}

In order to account for the large mass difference between the initial and final mesons, we follow Ref.~\cite{Yu:2018yxl} to introduce the reduction factor $\lambda $ existing in the off-diagonal terms:
\begin{eqnarray}
    \frac{1}{q^2-m_V^2+i\epsilon} \approx \frac{1}{(q^0)^2-m_V^2+i\epsilon} \approx -\lambda\frac{1}{m_V^2},
\end{eqnarray}
where $(q^0)^2=(\Delta M)^2$ and $\Delta M$ is the mass difference between the initial and final mesons. Since the $\eta_c$, $B_c^\pm$, and $\eta_b$ mesons are all quite massive, we have neglected their three-momenta to approximate the transferred momentum as $q^2 \approx (q^0)^2 = (\Delta M)^2$. Taking the process depicted in Fig.~\ref{fig-02}(a) as an example, numerically we obtain
\begin{eqnarray}
    \lambda_{\eta_c\eta_b \to B_c^+B_c^-} &=& \frac{-m_{B_c^*}^2}{(m_{\eta_b}-m_{B_c})^2-m_{B_c^*}^2}= 1.32.
\end{eqnarray}

\subsection{$VP$ interaction}
\label{sec:kernel-2}

\begin{figure}[h]
    \centering
    \includegraphics[width=.95\linewidth]{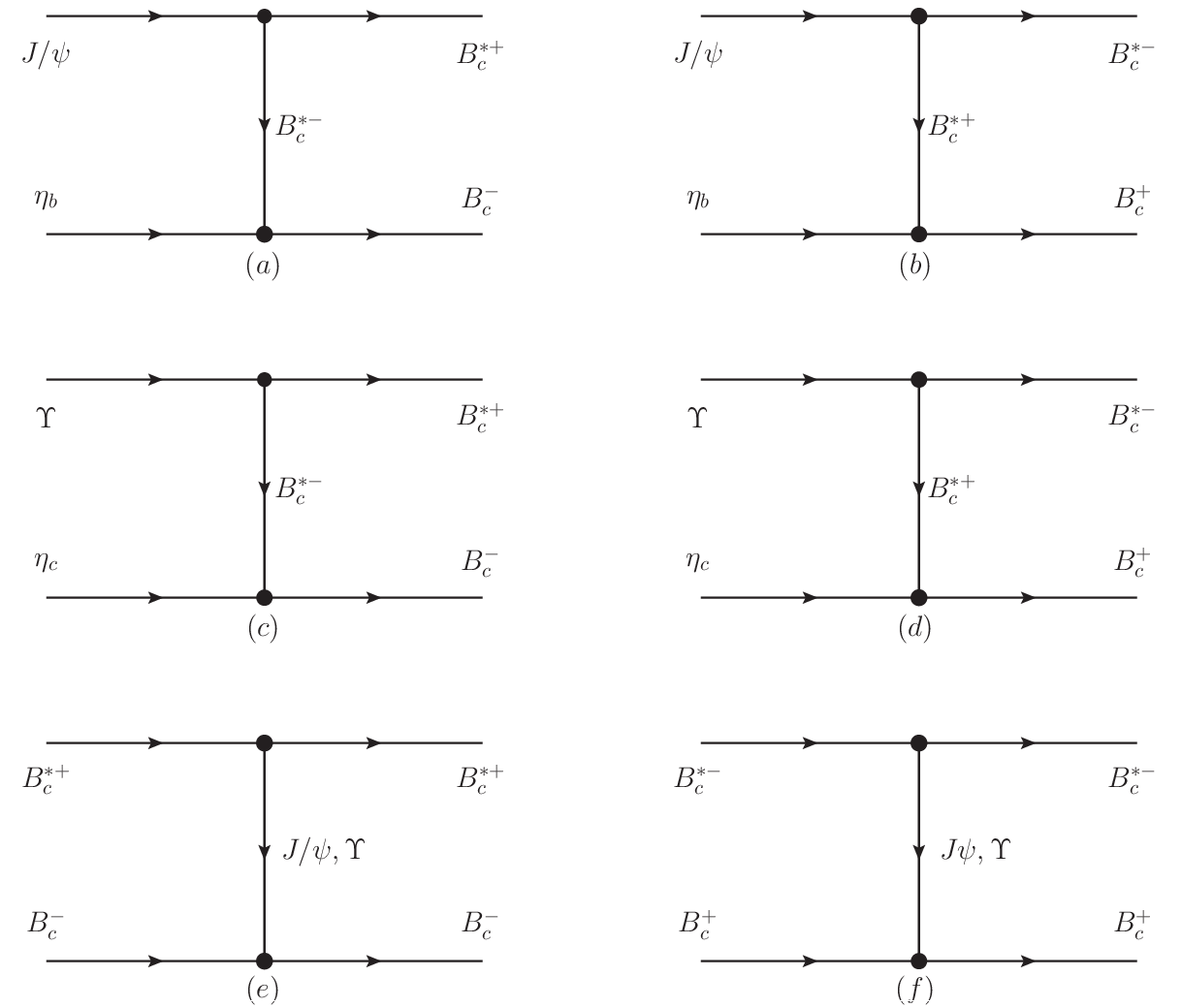}
    \caption{The vector meson exchange between vector and pseudoscalar mesons: (a) the $t$-channel diagram for $J/\psi\eta_b \to B_c^{*+}B_c^-$, (b) the $t$-channel diagram for $\Upsilon\eta_c \to B_c^{*+}B_c^-$, and (e) the $t$-channel diagram for $B_c^{*+}B_c^- \to B_c^{*+}B_c^-$. The subfigures (b,d,f) are their corresponding charge-conjugated diagrams.}
    \label{fig-03}
\end{figure}

In this subsection we investigate the $VP$ interaction arising from the vector meson exchange between vector and pseudoscalar mesons. There are four coupled channels
\begin{equation}
    \nonumber
    J/\psi\eta_b \, , ~~ \Upsilon\eta_c \, , ~~ B_c^{*+}B_c^- \, , ~~ B_c^{*-}B_c^+ \, ,
\end{equation}
which can be naturally separated into the single channel with the positive $C$-parity:
\begin{equation}
    \nonumber
    B_c^{*}\bar B_c^{(C=+)} \equiv \left(B_c^{*+}B_c^- + c.c. \right)/\sqrt{2}  \, ,
\end{equation}
and three coupled channels with the negative $C$-parity:
\begin{equation}
    \nonumber
    J/\psi\eta_b \, , ~ \Upsilon\eta_c \, , ~ B_c^{*}\bar B_c^{(C=-)} \equiv \left(B_c^{*+}B_c^- - c.c. \right)/\sqrt{2}  \, .
\end{equation}
Their interactions are shown in Fig.~\ref{fig-03}, and the transition potential can be derived from Eqs.~(\ref{eq-lagrangians}) as
\begin{eqnarray}
        \nonumber V_{VP(\pm)}(s) &=& C_{VP(\pm)}^t \times g^2 (p_1+p_3) (p_2+p_4) ~ \epsilon_1 \cdot \epsilon_3
        \\ \nonumber &+& C_{VP(\pm)}^u \times g^2 (p_1+p_4) (p_2+p_3) ~ \epsilon_1 \cdot \epsilon_3 \, ,
        \\
\end{eqnarray}
where the symbol $\pm$ denotes $C = \pm$, $p_1(p_3)$ is the four-momentum of the initial(final) vector meson, $p_2(p_4)$ is the four-momentum of the initial(final) pseudoscalar meson, and $\epsilon_1(\epsilon_3)$ is the polarization vector of the initial(final) vector meson.

For the single channel with $C=+$, the two coefficients are
\begin{eqnarray}
    C_{VP(+)}^t &=&  -(\frac{1}{m_{J/\psi}^2}+\frac{1}{m_{\Upsilon}^2}) ,
    \\
    C_{VP(+)}^u &=& 0.
\end{eqnarray}
For the three coupled channels with $C = -$, the two $3\times3$ matrices are $C_{VP(-)}^u = {\bf 0}_{3\times3}$ and
\begin{eqnarray}
    &&C_{VP(-)}^t =
    \\ \nonumber &&
    \renewcommand{\arraystretch}{1.5}
    \left(
    \begin{array}{c|ccc}
            J=1                                   & J/\psi\eta_b & \Upsilon\eta_c & B_c^{*}\bar B_c^{(C=-)}
            \\ \hline
            J/\psi\eta_b                          & 0            & 0              & \frac{\sqrt{2}\lambda}{m^2_{B_c^*}}                \\
            \Upsilon\eta_c                        & 0            & 0              & \frac{\sqrt{2}\lambda}{m^2_{B_c^*}}                \\
            B_c^{*}\bar B_c^{(C=-)}                & \frac{\sqrt{2}\lambda}{m^2_{B_c^*}}            & \frac{\sqrt{2}\lambda}{m^2_{B_c^*}}              & -(\frac{1}{m_{J/\psi}^2}+\frac{1}{m_{\Upsilon}^2})
        \end{array}
    \right)\, .
\end{eqnarray}
In the above expressions, we have approximated $\epsilon^0 \approx 0$ for the external vector mesons, because their three-momenta can be ignored compared to their masses when working at the thresholds. This non-relativistic approximation is also employed in Ref.~\cite{Molina:2008jw}. For a more detailed discussion on the validity of this approach, see Refs.~\cite{Gulmez:2016scm,Geng:2016pmf,Du:2018gyn}. Accordingly, the $VP \to VP$ transition potential is similar to the $PP \to PP$ transition potential, and we just need to add the extra factor $\epsilon_1 \cdot \epsilon_3$.

\subsection{$VV$ interaction}
\label{sec:kernel-3}

\begin{figure}[h]
    \centering
    \includegraphics[width=.9\linewidth]{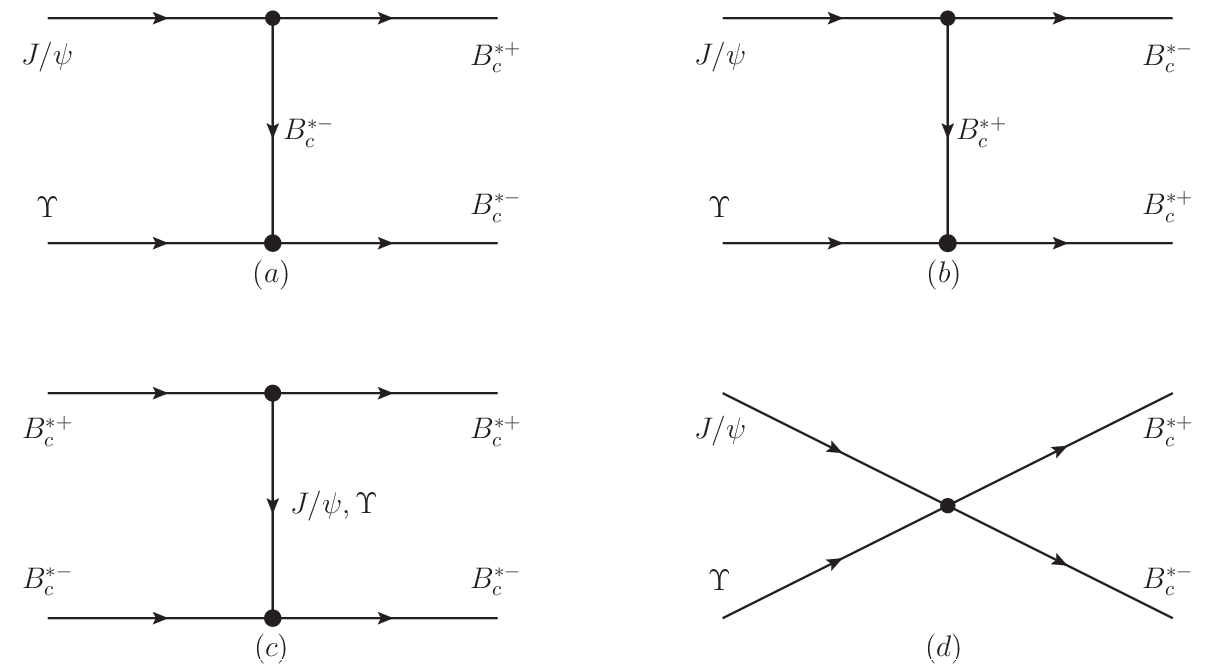}
    \caption{The vector meson exchange between two vector mesons: (a) the $t$-channel diagram for $J/\psi\Upsilon \to B_c^{*+}B_c^{*-}$, (b) the $u$-channel diagram for $J/\psi\Upsilon \to B_c^{*+}B_c^{*-}$, and (c) the $t$-channel diagram for $B_c^{*+}B_c^{*-} \to B_c^{*+}B_c^{*-}$. The subfigure (d) describes the contact term connecting four vector mesons, which also contributes here.}
    \label{fig-04}
\end{figure}

In this subsection we investigate the $VV$ interaction arising from the vector meson exchange between two vector mesons. There are two coupled channels:
\begin{equation}
    \nonumber
    J/\psi\Upsilon \, , ~~~ B_c^{*+}B_c^{*-} \, .
\end{equation}
Their interactions are shown in Fig.~\ref{fig-04}(a,b,c). Besides, we also need to consider the contact term shown in Fig.~\ref{fig-04}(d), so the transition potential consists of two terms
\begin{equation}
    V_{VV}(s)=V^{ex}_{VV}(s)+V^{co}_{VV}(s) \, .
\end{equation}

The term $V^{ex}_{VV}(s)$ describing the vector meson exchange can be derived from Eqs.~(\ref{eq-lagrangians}) as
\begin{eqnarray}
        \nonumber V^{ex}_{VV}(s) &=& C_{VV}^t \times g^2 (p_1+p_3) (p_2+p_4)\epsilon_1 \cdot \epsilon_3 \epsilon_2 \cdot \epsilon_4
        \\ \nonumber &+& C_{VV}^u \times g^2 (p_1+p_3) (p_2+p_4)\epsilon_1 \cdot \epsilon_4 \epsilon_2 \cdot \epsilon_3 \, ,
        \\
\end{eqnarray}
where $p_1(p_3)$ is the four-momentum of the $J/\psi(B_c^{*+})$ meson, $p_2(p_4)$ is the four-momentum of the $\Upsilon (B_c^{*-})$ meson, $\epsilon_1(\epsilon_3)$ is the polarization vector of the $J/\psi(B_c^{*+})$ meson, and $\epsilon_2(\epsilon_4)$ is the polarization vector of the $\Upsilon (B_c^{*-})$ meson. The two $2\times2$ matrices $C_{VV}^t$ and $C_{VV}^u$ are

\begin{eqnarray}
    \setlength{\arraycolsep}{5pt}
    \renewcommand{\arraystretch}{2}
    C_{VV}^t &=& \left(
    \begin{array}{c|cc}
            J=0,1,2              & J/\psi\Upsilon               & B_c^{*+}B_c^{*-}
            \\ \hline
            J/\psi\Upsilon   & 0                            & \lambda\frac{1}{m_{B_c^*}^2}                       \\
            B_c^{*+}B_c^{*-} & \lambda\frac{1}{m_{B_c^*}^2} & -(\frac{1}{m_{J/\psi}^2}+\frac{1}{m_{\Upsilon}^2})
        \end{array}
    \right)\, ,
\\
    \setlength{\arraycolsep}{5pt}
    \renewcommand{\arraystretch}{2}
    C_{VV}^u &=& \left(
    \begin{array}{c|cc}
            J=0,1,2              & J/\psi\Upsilon               & B_c^{*+}B_c^{*-}
            \\ \hline
            J/\psi\Upsilon   & 0                            & \lambda\frac{1}{m_{B_c^*}^2} \\
            B_c^{*+}B_c^{*-} & \lambda\frac{1}{m_{B_c^*}^2} & 0
        \end{array}
    \right)\, .
\end{eqnarray}

The contact term $V^{co}_{VV}(s)$ can be derived from Eqs.~(\ref{eq-lagrangians}) as:
\begin{eqnarray}
        && V^{co}_{J/\psi\Upsilon \to J/\psi\Upsilon}(s) = 0 \, ,\\
        \label{V-contact2}
        && V^{co}_{J/\psi\Upsilon \to B_c^{*+}B_c^{*-}}(s) \\
        \nonumber &=& g^2(-2\epsilon_\mu\epsilon^\mu\epsilon_\nu\epsilon^\nu + \epsilon_\mu\epsilon_\nu\epsilon^\mu\epsilon^\nu + \epsilon_\mu\epsilon_\nu\epsilon^\nu\epsilon^\mu) \, ,\\
        \label{V-contact3}
        && V^{co}_{B_c^{*+}B_c^{*-} \to B_c^{*+}B_c^{*-}}(s) \\
        \nonumber &=& g^2(2\epsilon_\mu\epsilon^\mu\epsilon_\nu\epsilon^\nu + 2\epsilon_\mu\epsilon_\nu\epsilon^\mu\epsilon^\nu - 4\epsilon_\mu\epsilon_\nu\epsilon^\nu\epsilon^\mu) \, .
\end{eqnarray}
As in previous treatments, we neglect the three-momentum of the external particles, which leads to the polarization vectors retaining only their spatial components. Consequently, a set of simplified spin projection operators can be constructed~\cite{Molina:2008jw}:
\begin{eqnarray}
    {\cal P}^{(0)}&=& \frac{1}{3}\epsilon_\mu \epsilon^\mu \epsilon_\nu \epsilon^\nu\, ,\\
    {\cal P}^{(1)}&=&\frac{1}{2}(\epsilon_\mu\epsilon_\nu\epsilon^\mu\epsilon^\nu-\epsilon_\mu\epsilon_\nu\epsilon^\nu\epsilon^\mu)\, ,\\
    {\cal P}^{(2)}&=&\frac{1}{2}(\epsilon_\mu\epsilon_\nu\epsilon^\mu\epsilon^\nu+\epsilon_\mu\epsilon_\nu\epsilon^\nu\epsilon^\mu)-\frac{1}{3}\epsilon_\mu\epsilon^\mu\epsilon_\nu\epsilon^\nu\, ,
\end{eqnarray}
to separate Eq.~(\ref{V-contact2}) and Eq.~(\ref{V-contact3}) into
\begin{eqnarray}
    V^{co}_{J/\psi\Upsilon \to B_c^{*+}B_c^{*-}}(s) &=&
    \left\{\begin{array}{cc}
        -4 g^2 & ~\textrm{for $J=0$}, \\
        0      & ~\textrm{for $J=1$}, \\
        2 g^2  & ~\textrm{for $J=2$},
    \end{array}\right.
\\
    V^{co}_{B_c^{*+}B_c^{*-} \to B_c^{*+}B_c^{*-}}(s) &=&
    \left\{\begin{array}{cc}
        4 g^2  & ~\textrm{for $J=0$}, \\
        6 g^2  & ~\textrm{for $J=1$}, \\
        -2 g^2 & ~\textrm{for $J=2$}.
    \end{array}\right.
\end{eqnarray}

\section{Results}
\label{sec:result}
In this letter we use the cutoff method to regularize the loop function $G(s)$ with the cutoff momentum $\Lambda$. This important parameter describes the dynamical scale to be integrated out, but its value is quite uncertain for the exchange of fully-heavy vector mesons. Hence, we choose a broad region, $\Lambda=400\sim1400\mev$, to perform numerical analyses. The resonances are dynamically generated as poles of the scattering amplitudes $T_{PP/VP/VV}(s)$. We find the possible existence of four poles in the fully-heavy $b \bar b c \bar c$ system that can qualify as hadronic molecules: one pole of $J^{PC}=0^{++}$ generated in the $PP$ interaction, two poles of $J^{PC}=1^{++}/1^{+-}$ generated in the $VP$ interaction, and one pole of $J^{PC}=2^{++}$ generated in the $VV$ interaction. Their positions are summarized in Table~\ref{tab:pole} with respect to the cutoff momentum $\Lambda$.

\begin{table*}[htb]
    \centering
    \caption{Pole positions $E_p$ with respect to the cutoff momentum $\Lambda$, in units of MeV. We only list the poles that can qualify as hadronic molecules.}
    \label{tab:pole}
    \setlength{\tabcolsep}{4.5pt}
    \renewcommand\arraystretch{1.4}
    \begin{tabular}{ccccccc}
        \hline
        \hline
        Pole                                            & $\Lambda = 400$     & $\Lambda = 600$    & $\Lambda = 800$   & $\Lambda = 1000$           & $\Lambda = 1200$       & $\Lambda = 1400$    \\
        \hline \hline
        $|B_c^{+} B_c^{-}; J^{PC}=0^{++}\rangle$        & $--$    & $12503.3-i126.8$               & $12383.1-i115.2$& $12379.0-i0$     & $12324.5-i0$ & $12227.8-i0$ \\
        \hline
        $|B_c^{*+} B_c^{-}; J^{PC}=1^{++}\rangle$       & $--$                & $--$       & $12604.8-i0$ & $12598.1-i0$ & $12581.5-i0$ & $12552.5-i0$ \\
        \hline
        $|B_c^{*+} B_c^{-}; J^{PC}=1^{+-}\rangle$       & $--$     & $12559.9-i111.6$   & $12495.8-i76.8$ & $12444.4-i42.2$    & $12394.8-i0$     & $12296.4-i0$ \\
        \hline
        $|B_c^{*+} B_c^{*-}; J^{PC}=0^{++}\rangle$      & $--$                & $--$               & $--$ & $--$ & $12549.4-i0$ & $12508.2-i0$ \\
        \hline
        $|B_c^{*+} B_c^{*-}; J^{PC}=1^{+-}\rangle$      & $--$                & $--$               & $--$ & $--$   & $--$   & $--$ \\
        \hline
        $|B_c^{*+} B_c^{*-}; J^{PC}=2^{++}\rangle$      & $12660.7-i45.7$       & $12572.0-i96.7$       & $12556.3-i0$& $12510.2-i0$  & $12418.8-i0$ & $12287.7-i0$ \\
        \hline
        \hline
    \end{tabular}
\end{table*}

\begin{figure}[hbtp]
\centering
\subfigbottomskip=2pt
\subfigcapskip=-1pt
\subfigure[]{\includegraphics[width=0.92\linewidth]{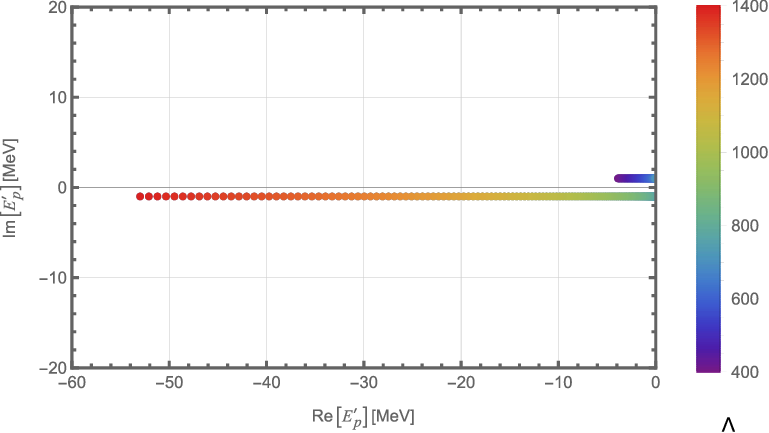}}
~~~~~
\subfigure[]{\includegraphics[width=0.92\linewidth]{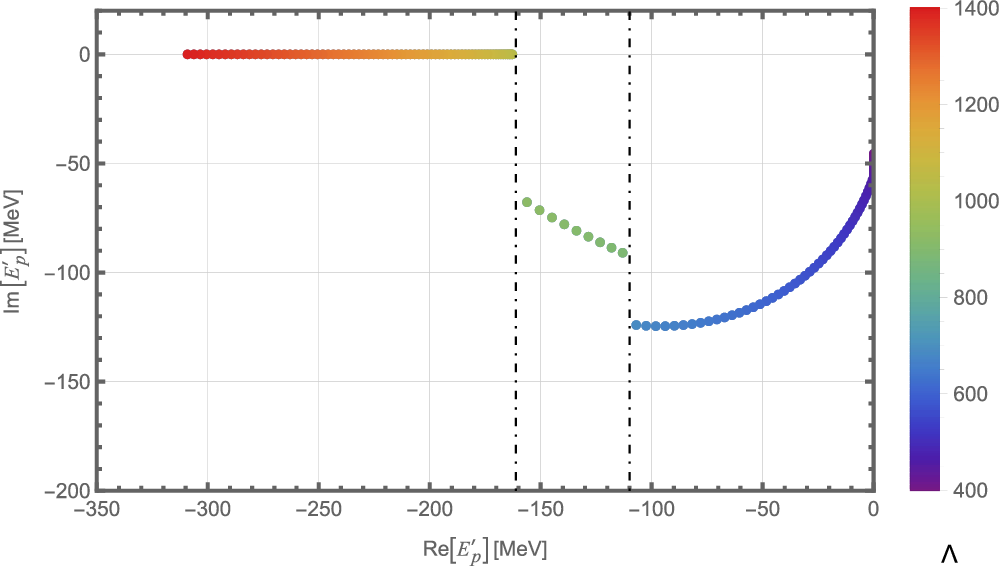}}
\caption{Pole positions $E_p^\prime = E_p - M_{B_c^*} - M_{B_c}$ of the hadronic molecules (a) $|B_c^{*+} B_c^{-}; J^{PC}=1^{++} \rangle$ and (b) $|B_c^{*+} B_c^{-}; J^{PC}=1^{+-} \rangle$ with respect to the cutoff momentum $\Lambda = 400 \sim 1400$~MeV. In subfigure (a), the first Riemann sheet (RI) and the second Riemann sheet (RII) are shown simultaneously, with a small imaginary part introduced to distinguish between them. In subfigure (b), the $J/\psi\eta_b$ and $\Upsilon\eta_c$ thresholds are indicated by dotted lines, and three different Riemann sheets are combined in the same plot for comparison: RI$(+++)$ is shown below the $\Upsilon\eta_c$ threshold, RII$(-++)$ appears between the $\Upsilon\eta_c$ and $J/\psi\eta_b$ thresholds, and RIII$(--+)$ is displayed above the $J/\psi\eta_b$ threshold.}
\label{fig:polepostion}
\end{figure}

Taking the second pole as an example, it appears in the $VP$ interaction with the positive $C$-parity and involves the single channel $B_c^{*}\bar B_c^{(C=+)}$, so we denote it as $|B_c^{*+} B_c^{-}; J^{PC}=1^{++} \rangle$. As shown in Fig.~\ref{fig:polepostion}(a), this pole is a virtual state on the second Riemann sheet when taking $\Lambda<710$~MeV, and it becomes a bound state on the first Riemann sheet when taking $\Lambda>710$~MeV. It the limit $M_{\Upsilon} \rightarrow +\infty $, only the exchange of the $J/\psi$ meson contributes, and this pole can still be a bound state when taking $\Lambda>940$~MeV. Therefore, our results suggest the possible existence of the hadronic molecule $|B_c^{*+} B_c^{-}; J^{PC}=1^{++} \rangle$ for $\Lambda>710$~MeV, whose interaction mainly arises from the exchange of the $J/\psi$ meson.

Taking the third pole as another example, it appears in the $VP$ interaction with the negative $C$-parity and involves three coupled channels $J/\psi\eta_b$, $\Upsilon\eta_c$, and $B_c^{*}\bar B_c^{(C=-)}$. As shown in Fig.\ref{fig:polepostion}(b), three distinct poles are present. However, under a specific cutoff momentum $\Lambda$, only one of these poles appears near the physical sheet, while the others lie on Riemann sheets farther from the physical region. Accordingly, we group together the poles that are close to the physical sheet and display them collectively in Fig.\ref{fig:polepostion}(b). In order to describe the coupling strength of this pole to the three channels, we introduce the coupling parameters $g_i$. They are defined in the vicinity of the pole as
\begin{equation}
    \label{eq:gi1}
    T_{PP/VP/VV}^{ij}(s) = \frac{g_ig_j}{s-E_p^2} \, ,
\end{equation}
with $E_p$ the pole position. We take $\Lambda = 500$~MeV and summarize the detailed pole information in Table~\ref{tab:result}. The third pole strongly couples to the $B_c^{*} \bar B_c^{(C=-)}$ channel, so we denote it as $|B_c^{*+} B_c^{-}; J^{PC}=1^{+-} \rangle$. Therefore, our results suggest the existence of this hadronic molecule for $\Lambda>460$~MeV, where the coupled-channel effects are important. Besides, our results suggest the existence of the hadronic molecules $|B_c^{+} B_c^{-}; J^{PC}=0^{++} \rangle$ and $|B_c^{*+} B_c^{*-}; J^{PC}=2^{++} \rangle$, whose pole information is also summarized in Table~\ref{tab:result}. Oppositely, the hadronic molecules $|B_c^{*+} B_c^{*-}; J^{PC}=0^{++} \rangle$ and $|B_c^{*+} B_c^{*-}; J^{PC}=1^{+-} \rangle$ are less likely to exist within our approach.

\begin{table}[h!]
\centering
\renewcommand{\arraystretch}{1.6}
\caption{Pole positions $E_p$ and their couplings $g_i$ to various coupled channels, with the cutoff momentum $\Lambda=500\mev$. We only list the poles that can qualify as hadronic molecules.}
\small 
\begin{tabular}{p{3.2cm}|p{1.5cm}|p{1.4cm}|p{0.6cm}} 
\hline\hline
State & $E_p$ (MeV) & Channel & $|g_i|$ (GeV) \\ \hline\hline
\multirow{2}{*}{$|B_c^{+} B_c^{-}; J^{PC}=0^{++}\rangle$} & \multirow{2}{*}{$12541-i83$} & $\eta_c\eta_b$ & 20 \\ \cline{3-4}
 & & $B_c^+B_c^-$ & 123 \\ \hline
\multirow{3}{*}{$|B_c^{*+} B_c^{-}; J^{PC}=1^{+-}\rangle$} & \multirow{3}{*}{$12597-i74$} & $J/\psi\eta_b$ & 14 \\ \cline{3-4}
 & & $\Upsilon\eta_c$ & 14 \\ \cline{3-4}
 & & $B_c^{*} \bar B_c^{(C=-)}$ & 114 \\ \hline
\multirow{2}{*}{$|B_c^{*+} B_c^{*-}; J^{PC}=2^{++}\rangle$} & \multirow{2}{*}{$12630-i80$} & $J/\psi\Upsilon$ & 23 \\ \cline{3-4}
 & & $B_c^{*+}B_c^{*-}$ & 135 \\ \hline\hline
\end{tabular}
\label{tab:result}
\end{table}

\begin{figure}[h]
\centering
\includegraphics[width=.92\linewidth]{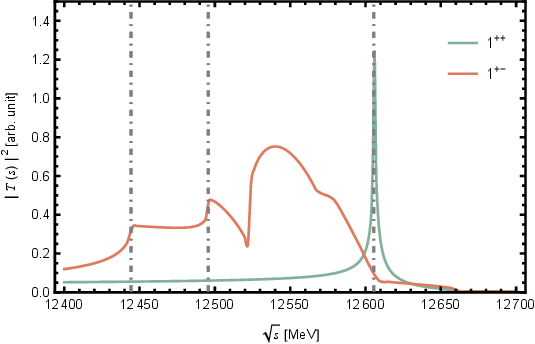}
\caption{The scattering amplitudes $|T_{VP}^{B_c^{*} \bar B_c^{(C=+)} \to B_c^{*} \bar B_c^{(C=+)}}(s)|^2$ and $|T_{VP}^{B_c^{*} \bar B_c^{(C=-)} \to B_c^{*} \bar B_c^{(C=-)}}(s)|^2$ for the cutoff momentum $\Lambda = 600$~MeV.}
\label{fig:scattering}
\end{figure}

It is interesting to further study the pole properties. We still take the second and third poles as examples and show their scattering amplitudes $T_{VP}^{B_c^{*} \bar B_c^{(C=+)} \to B_c^{*} \bar B_c^{(C=+)}}(s)$ and $T_{VP}^{B_c^{*} \bar B_c^{(C=-)} \to B_c^{*} \bar B_c^{(C=-)}}(s)$ in Fig.~\ref{fig:scattering} for the cutoff momentum $\Lambda = 600$~MeV. The second pole of $J^{PC}=1^{++}$ is identified as a virtual state, and it enhances the near-threshold cusp effect to produce a sharp peak at the $B_c^{*+} B_c^{-}$ threshold. Differently, the third pole of $J^{PC}=1^{+-}$ is identified as a bound state, and it appears as a normal peak under the $B_c^{*+} B_c^{-}$ threshold, but there also exist the threshold effects. Detailed discussions on the near-threshold virtual state can be found in Ref.~\cite{Dong:2020hxe}.

\section{Conclusions}
\label{sec:conclu}
We apply the coupled-channel unitary approach within the local hidden-gauge formalism to study the fully-heavy system $b \bar b c \bar c$, and the obtained results suggest the possible existence of the fully-heavy hadronic molecules $B_c^{(*)+} B_c^{(*)-}$, whose interactions mainly arise from the exchange of the fully-heavy vector mesons $J/\psi$, $B_c^*$, and $\Upsilon$. Especially, $|B_c^{+} B_c^{-}; J^{PC}=0^{++} \rangle$, $|B_c^{*+} B_c^{-}; J^{PC}=1^{+-} \rangle$, and $|B_c^{*+} B_c^{*-}; J^{PC}=2^{++} \rangle$ are more likely to exist, while $|B_c^{*+} B_c^{-}; J^{PC}=1^{++} \rangle$ may exist or it may also behave as a threshold cusp. We do not take into account the direct gluon exchange to avoid the possible double counting, but this interaction can be attractive after properly arranging the relative color structures, so the above conclusion remains. Besides, we have investigated the fully-heavy systems $b c \bar c \bar c$,  $b b \bar c \bar c$, and $b b \bar b \bar c$, but our results do not support the existence of deeply-bound hadronic molecules in these systems.

Compared to the exchange of light mesons, the exchange of fully-heavy mesons are much less understood, and it is still controversial whether the induced interaction can be large enough to form hadronic molecules. Accordingly, the fully-heavy hadronic molecules $B_c^{(*)+} B_c^{(*)-}$ predicted in this letter are of particular interest, since their studies can be taken as a general investigation on the slightly different question whether a lower-mass fully-heavy meson is capable of binding two higher-mass fully-heavy hadrons. We propose to search for these potential exotic hadrons in the $J/\psi \Upsilon$, $\mu^+ \mu^- J/\psi$, and $\mu^+ \mu^- \Upsilon$ channels in future ATLAS, CMS, and LHCb experiments.

\begin{acknowledgements}
The authors are grateful to Eulogio Oset for sharing his insights into the considered topics and for the very helpful discussions.
This project is supported by
the National Natural Science Foundation of China under Grant No.~12075019,
the Jiangsu Provincial Double-Innovation Program under Grant No.~JSSCRC2021488,
and
the Fundamental Research Funds for the Central Universities.
\end{acknowledgements}


\begin{thebibliography}{99}


\bibitem{LHCb:2015yax}
R.~Aaij \textit{et al.} [LHCb],
{Observation of $J/\psi p$ Resonances Consistent with Pentaquark States in $\Lambda_b^0 \to J/\psi K^- p$ Decays,}
Phys. Rev. Lett. \textbf{115}, 072001 (2015).
\href{https://doi.org/10.1103/PhysRevLett.115.072001}{\path{doi:10.1103/PhysRevLett.115.072001}},
\href{https://arXiv.org/abs/1507.03414}{\path{arXiv:1507.03414}}.


\bibitem{LHCb:2019kea}
R.~Aaij \textit{et al.} [LHCb],
{Observation of a narrow pentaquark state, $P_c(4312)^+$, and of two-peak structure of the $P_c(4450)^+$,}
Phys. Rev. Lett. \textbf{122}, no.22, 222001 (2019).
\href{https://doi.org/10.1103/PhysRevLett.122.222001}{\path{doi:10.1103/PhysRevLett.122.222001}},
\href{https://arXiv.org/abs/1904.03947}{\path{arXiv:1904.03947}}.

\bibitem{LHCb:2020jpq}
R.~Aaij \textit{et al.} [LHCb],
{Evidence of a $J/\psi\Lambda$ structure and observation of excited $\Xi^-$ states in the $\Xi^-_b \to J/\psi\Lambda K^-$ decay,}
Sci. Bull. \textbf{66}, 1278-1287 (2021).
\href{https://doi.org/10.1016/j.scib.2021.02.030}{\path{doi:10.1016/j.scib.2021.02.030}},
\href{https://arXiv.org/abs/2012.10380}{\path{arXiv:2012.10380}}.

\bibitem{LHCb:2022ogu}
R.~Aaij \textit{et al.} [LHCb],
{Observation of a J/\ensuremath{\psi}\ensuremath{\Lambda} Resonance Consistent with a Strange Pentaquark Candidate in B-\textrightarrow{}J/\ensuremath{\psi}\ensuremath{\Lambda}p\textasciimacron{} Decays,}
Phys. Rev. Lett. \textbf{131}, no.3, 031901 (2023).
\href{https://doi.org/10.1103/PhysRevLett.131.031901}{\path{doi:10.1103/PhysRevLett.131.031901}},
\href{https://arXiv.org/abs/2210.10346}{\path{arXiv:2210.10346}}.

\bibitem{Wu:2010jy}
J.~J.~Wu, R.~Molina, E.~Oset and B.~S.~Zou,
{Prediction of narrow $N^*$ and $\Lambda^*$ resonances with hidden charm above 4 GeV,}
Phys. Rev. Lett. \textbf{105}, 232001 (2010).
\href{https://doi.org/10.1103/PhysRevLett.105.232001}{\path{doi:10.1103/PhysRevLett.105.232001}},
\href{https://arXiv.org/abs/1007.0573}{\path{arXiv:1007.0573}}.

\bibitem{Chen:2015sxa}
H.~X.~Chen, L.~S.~Geng, W.~H.~Liang, E.~Oset, E.~Wang and J.~J.~Xie,
{Looking for a hidden-charm pentaquark state with strangeness $S=-1$ from $\Xi^-_b$ decay into $J/\psi K^- \Lambda$,}
Phys. Rev. C \textbf{93} (2016) no.6, 065203.
\href{https://doi.org/10.1103/PhysRevC.93.065203}{\path{doi:10.1103/PhysRevC.93.065203}},
\href{https://arXiv.org/abs/1510.01803}{\path{arXiv:1510.01803}}.

\bibitem{Xiao:2019gjd}
C.~W.~Xiao, J.~Nieves and E.~Oset,
{Prediction of hidden charm strange molecular baryon states with heavy quark spin symmetry,}
Phys. Lett. B \textbf{799}, 135051 (2019).
\href{https://doi.org/10.1016/j.physletb.2019.135051}{\path{doi:10.1016/j.physletb.2019.135051}},
\href{https://arXiv.org/abs/1906.09010}{\path{arXiv:1906.09010}}.

\bibitem{Wang:2011rga}
W.~L.~Wang, F.~Huang, Z.~Y.~Zhang and B.~S.~Zou,
{$\Sigma_c \bar{D}$ and $\Lambda_c \bar{D}$ states in a chiral quark model,}
Phys. Rev. C \textbf{84}, 015203 (2011).
\href{https://doi.org/10.1103/PhysRevC.84.015203}{\path{doi:10.1103/PhysRevC.84.015203}},
\href{https://arXiv.org/abs/1101.0453}{\path{arXiv:1101.0453}}.

\bibitem{Yang:2011wz}
Z.~C.~Yang, Z.~F.~Sun, J.~He, X.~Liu and S.~L.~Zhu,
{The possible hidden-charm molecular baryons composed of anti-charmed meson and charmed baryon,}
Chin. Phys. C \textbf{36}, 6-13 (2012).
\href{https://doi.org/10.1088/1674-1137/36/1/002}{\path{doi:10.1088/1674-1137/36/1/002}},
\href{https://arXiv.org/abs/1105.2901}{\path{arXiv:1105.2901}}.

\bibitem{Karliner:2015ina}
M.~Karliner and J.~L.~Rosner,
{New Exotic Meson and Baryon Resonances from Doubly-Heavy Hadronic Molecules,}
Phys. Rev. Lett. \textbf{115}, no.12, 122001 (2015).
\href{https://doi.org/10.1103/PhysRevLett.115.122001}{\path{doi:10.1103/PhysRevLett.115.122001}},
\href{https://arXiv.org/abs/1506.06386}{\path{arXiv:1506.06386}}.

\bibitem{Cheng:2015cca}
H.~Y.~Cheng and C.~K.~Chua,
{Bottom Baryon Decays to Pseudoscalar Meson and Pentaquark,}
Phys. Rev. D \textbf{92}, no.9, 096009 (2015).
\href{https://doi.org/10.1103/PhysRevD.92.096009}{\path{doi:10.1103/PhysRevD.92.096009}},
\href{https://arXiv.org/abs/1509.03708}{\path{arXiv:1509.03708}}.

\bibitem{Santopinto:2016pkp}
E.~Santopinto and A.~Giachino,
{Compact pentaquark structures,}
Phys. Rev. D \textbf{96}, no.1, 014014 (2017).
\href{https://doi.org/10.1103/PhysRevD.96.014014}{\path{doi:10.1103/PhysRevD.96.014014}},
\href{https://arXiv.org/abs/1604.03769}{\path{arXiv:1604.03769}}.

\bibitem{Chen:2016ryt}
R.~Chen, J.~He and X.~Liu,
{Possible strange hidden-charm pentaquarks from $\Sigma_c^{(*)}\bar{D}_s^*$ and $\Xi^{(',*)}_c\bar{D}^*$ interactions,}
Chin. Phys. C \textbf{41}, no.10, 103105 (2017).
\href{https://doi.org/10.1088/1674-1137/41/10/103105}{\path{doi:10.1088/1674-1137/41/10/103105}},
\href{https://arXiv.org/abs/1609.03235}{\path{arXiv:1609.03235}}.

\bibitem{Shen:2019evi}
C.~W.~Shen, J.~J.~Wu and B.~S.~Zou,
{Decay behaviors of possible $\Lambda_{c\bar{c}}$ states in hadronic molecule pictures,}
Phys. Rev. D \textbf{100}, no.5, 056006 (2019).
\href{https://doi.org/10.1103/PhysRevD.100.056006}{\path{doi:10.1103/PhysRevD.100.056006}},
\href{https://arXiv.org/abs/1906.03896}{\path{arXiv:1906.03896}}.

\bibitem{Wang:2019nvm}
B.~Wang, L.~Meng and S.~L.~Zhu,
{Spectrum of the strange hidden charm molecular pentaquarks in chiral effective field theory,}
Phys. Rev. D \textbf{101}, no.3, 034018 (2020).
\href{https://doi.org/10.1103/PhysRevD.101.034018}{\path{doi:10.1103/PhysRevD.101.034018}},
\href{https://arXiv.org/abs/1912.12592}{\path{arXiv:1912.12592}}.

\bibitem{Shen:2020gpw}
C.~W.~Shen, H.~J.~Jing, F.~K.~Guo and J.~J.~Wu,
{Exploring Possible Triangle Singularities in the $\Xi^-_{b} \to K^- J/\psi \Lambda$ Decay,}
Symmetry \textbf{12}, no.10, 1611 (2020).
\href{https://doi.org/10.3390/sym12101611}{\path{doi:10.3390/sym12101611}},
\href{https://arXiv.org/abs/2008.09082}{\path{arXiv:2008.09082}}.

\bibitem{Oller:1997ti}
J.~A.~Oller and E.~Oset,
{Chiral symmetry amplitudes in the S wave isoscalar and isovector channels and the $\sigma$, f$_0$(980), a$_0$(980) scalar mesons,}
Nucl. Phys. A \textbf{620}, 438-456 (1997).
\href{https://doi.org/10.1016/S0375-9474(97)00160-7}{\path{doi:10.1016/S0375-9474(97)00160-7}},
\href{https://arXiv.org/pdf/hep-ph/9702314}{\path{arXiv:hep-ph/9702314}}.

\bibitem{Oller:1998zr}
J.~A.~Oller and E.~Oset,
{N/D description of two meson amplitudes and chiral symmetry,}
Phys. Rev. D \textbf{60}, 074023 (1999).
\href{https://doi.org/10.1103/PhysRevD.60.074023}{\path{doi:10.1103/PhysRevD.60.074023}},
\href{https://arXiv.org/pdf/hep-ph/9809337}{\path{arXiv:hep-ph/9809337}}.

\bibitem{GomezNicola:2001as}
A.~Gomez Nicola and J.~R.~Pelaez,
{Meson meson scattering within one loop chiral perturbation theory and its unitarization,}
Phys. Rev. D \textbf{65}, 054009 (2002).
\href{https://doi.org/10.1103/PhysRevD.65.054009}{\path{doi:10.1103/PhysRevD.65.054009}},
\href{https://arXiv.org/pdf/hep-ph/0109056}{\path{arXiv:hep-ph/0109056}}.





\bibitem{Oller:2000fj}
J.~A.~Oller and U.~G.~Meissner,
{Chiral dynamics in the presence of bound states: Kaon nucleon interactions revisited,}
Phys. Lett. B \textbf{500}, 263-272 (2001).
\href{https://doi.org/doi:10.1016/S0370-2693(01)00078-8}{\path{doi:10.1016/S0370-2693(01)00078-8}},
\href{https://arXiv.org/pdf/hep-ph/0011146}{\path{arXiv:hep-ph/0011146}}.

\bibitem{Jido:2003cb}
D.~Jido, J.~A.~Oller, E.~Oset, A.~Ramos and U.~G.~Meissner,
{Chiral dynamics of the two Lambda(1405) states,}
Nucl. Phys. A \textbf{725}, 181-200 (2003).
\href{https://doi.org/10.1016/S0375-9474(03)01598-7}{\path{doi:10.1016/S0375-9474(03)01598-7}},
\href{https://arXiv.org/pdf/nucl-th/0303062}{\path{arXiv:nucl-th/0303062}}.

\bibitem{Hofmann:2003je}
J.~Hofmann and M.~F.~M.~Lutz,
{Open charm meson resonances with negative strangeness,}
Nucl. Phys. A \textbf{733}, 142-152 (2004).
\href{https://doi.org/10.1016/j.nuclphysa.2003.12.013}{\path{doi:10.1016/j.nuclphysa.2003.12.013}},
\href{https://arXiv.org/pdf/hep-ph/0308263}{\path{arXiv:hep-ph/0308263}}.

\bibitem{Geng:2008gx}
L.~S.~Geng and E.~Oset,
{Vector meson-vector meson interaction in a hidden gauge unitary approach,}
Phys. Rev. D \textbf{79}, 074009 (2009).
\href{https://doi.org/10.1103/PhysRevD.79.074009}{\path{doi:10.1103/PhysRevD.79.074009}},
\href{https://arXiv.org/abs/0812.1199}{\path{arXiv:0812.1199}}.

\bibitem{Guo:2006fu}
F.~K.~Guo, P.~N.~Shen, H.~C.~Chiang, R.~G.~Ping and B.~S.~Zou,
{Dynamically generated 0+ heavy mesons in a heavy chiral unitary approach,}
Phys. Lett. B \textbf{641}, 278-285 (2006).
\href{https://doi.org/10.1016/j.physletb.2006.08.064}{\path{doi:10.1016/j.physletb.2006.08.064}},
\href{https://arXiv.org/pdf/hep-ph/0603072}{\path{arXiv:hep-ph/0603072}}.

\bibitem{Mizutani:2006vq}
T.~Mizutani and A.~Ramos,
{D mesons in nuclear matter: A DN coupled-channel equations approach,}
Phys. Rev. C \textbf{74}, 065201 (2006).
\href{https://doi.org/10.1103/PhysRevC.74.065201}{\path{doi:10.1103/PhysRevC.74.065201}},
\href{https://arXiv.org/pdf/hep-ph/0607257}{\path{arXiv:hep-ph/0607257}}.

\bibitem{Lu:2016gev}
J.~X.~Lu, H.~X.~Chen, Z.~H.~Guo, J.~Nieves, J.~J.~Xie and L.~S.~Geng,
{$\Lambda_c(2595)$ resonance as a dynamically generated state: The compositeness condition and the large $N_c$ evolution,}
Phys. Rev. D \textbf{93}, no.11, 114028 (2016).
\href{https://doi.org/10.1103/PhysRevD.93.114028}{\path{doi:10.1103/PhysRevD.93.114028}},
\href{https://arXiv.org/abs/1603.05388}{\path{arXiv:1603.05388}}.






\bibitem{Chen:2022asf}
H.~X.~Chen, W.~Chen, X.~Liu, Y.~R.~Liu and S.~L.~Zhu,
{An updated review of the new hadron states,}
Rept. Prog. Phys. \textbf{86}, no.2, 026201 (2023).
\href{https://doi.org/10.1088/1361-6633/aca3b6}{\path{doi:10.1088/1361-6633/aca3b6}},
\href{https://arXiv.org/abs/2204.02649}{\path{arXiv:2204.02649}}.

\bibitem{Guo:2017jvc}
F.~K.~Guo, C.~Hanhart, U.~G.~Mei\ss{}ner, Q.~Wang, Q.~Zhao and B.~S.~Zou,
{Hadronic molecules,}
Rev. Mod. Phys. \textbf{90}, no.1, 015004 (2018).
[erratum: Rev. Mod. Phys. \textbf{94}, no.2, 029901 (2022)].
\href{https://doi.org/10.1103/RevModPhys.90.015004}{\path{doi:10.1103/RevModPhys.90.015004}},
\href{https://arXiv.org/abs/1705.00141 }{\path{arXiv:1705.00141 }}.

\bibitem{LHCb:2020bwg}
R.~Aaij \textit{et al.} [LHCb],
{Observation of structure in the $J /\psi$ -pair mass spectrum,}
Sci. Bull. \textbf{65}, no.23, 1983-1993 (2020).
\href{https://doi.org/10.1016/j.scib.2020.08.032}{\path{doi:10.1016/j.scib.2020.08.032}},
\href{https://arXiv.org/abs/2006.16957}{\path{arXiv:2006.16957}}.

\bibitem{ATLAS:2023bft}
G.~Aad \textit{et al.} [ATLAS],
{Observation of an Excess of Dicharmonium Events in the Four-Muon Final State with the ATLAS Detector,}
Phys. Rev. Lett. \textbf{131}, no.15, 151902 (2023).
\href{https://doi.org/10.1103/PhysRevLett.131.151902}{\path{doi:10.1103/PhysRevLett.131.151902}},
\href{https://arXiv.org/abs/2304.08962}{\path{arXiv:2304.08962}}.

\bibitem{CMS:2023owd}
A.~Hayrapetyan \textit{et al.} [CMS],
{New Structures in the J/\ensuremath{\psi}J/\ensuremath{\psi} Mass Spectrum in Proton-Proton Collisions at s=13\,\,TeV,}
Phys. Rev. Lett. \textbf{132}, no.11, 111901 (2024).
\href{https://doi.org/10.1103/PhysRevLett.132.111901}{\path{doi:10.1103/PhysRevLett.132.111901}},
\href{https://arXiv.org/abs/2306.07164}{\path{arXiv:2306.07164}}.

\bibitem{Chen:2016jxd}
W.~Chen, H.~X.~Chen, X.~Liu, T.~G.~Steele and S.~L.~Zhu,
{Hunting for exotic doubly hidden-charm/bottom tetraquark states,}
Phys. Lett. B \textbf{773}, 247-251 (2017).
\href{https://doi.org/10.1016/j.physletb.2017.08.034}{\path{doi:10.1016/j.physletb.2017.08.034}},
\href{https://arXiv.org/abs/1605.01647}{\path{arXiv:1605.01647}}.

\bibitem{Czarnecki:2017vco}
A.~Czarnecki, B.~Leng and M.~B.~Voloshin,
{Stability of tetrons,}
Phys. Lett. B \textbf{778}, 233-238 (2018).
\href{https://doi.org/10.1016/j.physletb.2018.01.034}{\path{doi:10.1016/j.physletb.2018.01.034}},
\href{https://arXiv.org/abs/1708.04594}{\path{arXiv:1708.04594}}.

\bibitem{Guo:2020pvt}
Z.~H.~Guo and J.~A.~Oller,
{Insights into the inner structures of the fully charmed tetraquark state $X(6900)$,}
Phys. Rev. D \textbf{103}, no.3, 034024 (2021).
\href{https://doi.org/10.1103/PhysRevD.103.034024}{\path{doi:10.1103/PhysRevD.103.034024}},
\href{https://arXiv.org/abs/2011.00978 }{\path{arXiv:2011.00978 }}.

\bibitem{Cao:2020gul}
Q.~F.~Cao, H.~Chen, H.~R.~Qi and H.~Q.~Zheng,
{Some remarks on X(6900),}
Chin. Phys. C \textbf{45}, no.10, 103102 (2021).
\href{https://doi.org/10.1088/1674-1137/ac0ee5}{\path{doi:10.1088/1674-1137/ac0ee5}},
\href{https://arXiv.org/abs/2011.04347}{\path{arXiv:2011.04347}}.

\bibitem{Gong:2020bmg}
C.~Gong, M.~C.~Du, Q.~Zhao, X.~H.~Zhong and B.~Zhou,
{Nature of X(6900) and its production mechanism at LHCb,}
Phys. Lett. B \textbf{824}, 136794 (2022).
\href{https://doi.org/10.1016/j.physletb.2021.136794}{\path{doi:10.1016/j.physletb.2021.136794}},
\href{https://arXiv.org/abs/2011.11374}{\path{arXiv:2011.11374}}.

\bibitem{Dong:2021lkh}
X.~K.~Dong, V.~Baru, F.~K.~Guo, C.~Hanhart, A.~Nefediev and B.~S.~Zou,
{Is the existence of a J/\ensuremath{\psi}J/\ensuremath{\psi} bound state plausible?,}
Sci. Bull. \textbf{66}, no.24, 2462-2470 (2021).
\href{https://doi.org/10.1016/j.scib.2021.09.009}{\path{doi:10.1016/j.scib.2021.09.009}},
\href{https://arXiv.org/abs/2107.03946}{\path{arXiv:2107.03946}}.

\bibitem{Wu:2016vtq}
J.~Wu, Y.~R.~Liu, K.~Chen, X.~Liu and S.~L.~Zhu,
{Heavy-flavored tetraquark states with the $QQ\bar{Q}\bar{Q}$ configuration,}
Phys. Rev. D \textbf{97}, no.9, 094015 (2018).
\href{https://doi.org/10.1103/PhysRevD.97.094015}{\path{doi:10.1103/PhysRevD.97.094015}},
\href{https://arXiv.org/abs/1605.01134}{\path{arXiv:1605.01134}}.

\bibitem{Richard:2017vry}
J.~M.~Richard, A.~Valcarce and J.~Vijande,
{String dynamics and metastability of all-heavy tetraquarks,}
Phys. Rev. D \textbf{95}, no.5, 054019 (2017).
\href{https://doi.org/10.1103/PhysRevD.95.054019}{\path{doi:10.1103/PhysRevD.95.054019}},
\href{https://arXiv.org/abs/1703.00783}{\path{arXiv:1703.00783}}.

\bibitem{Anwar:2017toa}
M.~N.~Anwar, J.~Ferretti, F.~K.~Guo, E.~Santopinto and B.~S.~Zou,
{Spectroscopy and decays of the fully-heavy tetraquarks,}
Eur. Phys. J. C \textbf{78}, no.8, 647 (2018).
\href{https://doi.org/10.1140/epjc/s10052-018-6073-9}{\path{doi:10.1140/epjc/s10052-018-6073-9}},
\href{https://arXiv.org/abs/1710.02540}{\path{arXiv:1710.02540}}.

\bibitem{Yang:2021zrc}
Z.~H.~Yang, Q.~N.~Wang, W.~Chen and H.~X.~Chen,
{Investigation of the stability for fully-heavy $bc\bar{b}\bar{c}$ tetraquark states,}
Phys. Rev. D \textbf{104}, no.1, 014003 (2021).
\href{https://doi.org/10.1103/PhysRevD.104.014003}{\path{doi:10.1103/PhysRevD.104.014003}},
\href{https://arXiv.org/abs/2102.10605}{\path{arXiv:2102.10605}}.

\bibitem{Wang:2021mma}
Q.~N.~Wang, Z.~Y.~Yang and W.~Chen,
{Exotic fully-heavy $Q\bar QQ\bar Q$ tetraquark states in $\mathbf{8}_{[Q\bar{Q}]}\otimes \mathbf{8}_{[Q\bar{Q}]}$ color configuration,}
Phys. Rev. D \textbf{104}, no.11, 114037 (2021).
\href{https://doi.org/10.1103/PhysRevD.104.114037}{\path{doi:10.1103/PhysRevD.104.114037}},
\href{https://arXiv.org/abs/2109.08091}{\path{arXiv:2109.08091}}.

\bibitem{Zhang:2022qtp}
J.~Zhang, J.~B.~Wang, G.~Li, C.~S.~An, C.~R.~Deng and J.~J.~Xie,
{Spectrum of the S-wave fully-heavy tetraquark states,}
Eur. Phys. J. C \textbf{82}, no.12, 1126 (2022).
\href{https://doi.org/10.1140/epjc/s10052-022-11111-4}{\path{doi:10.1140/epjc/s10052-022-11111-4}},
\href{https://arXiv.org/abs/2209.13856}{\path{arXiv:2209.13856}}.

\bibitem{Bando:1987br}
M.~Bando, T.~Kugo and K.~Yamawaki,
{Nonlinear Realization and Hidden Local Symmetries,}
Phys. Rept. \textbf{164}, 217-314 (1988).
\href{https://doi.org/10.1016/0370-1573(88)90019-1}{\path{doi:10.1016/0370-1573(88)90019-1}}.

\bibitem{Meissner:1987ge}
U.~G.~Meissner,
{Low-Energy Hadron Physics from Effective Chiral Lagrangians with Vector Mesons,}
Phys. Rept. \textbf{161}, 213 (1988).
\href{https://doi.org/10.1016/0370-1573(88)90090-7}{doi:\path{10.1016/0370-1573(88)90090-7}}.

\bibitem{Oset:2010tof}
E.~Oset and A.~Ramos,
{Dynamically generated resonances from the vector octet-baryon octet interaction,}
Eur. Phys. J. A \textbf{44}, 445-454 (2010).
\href{https://doi.org/10.1140/epja/i2010-10957-3}{\path{doi:10.1140/epja/i2010-10957-3}},
\href{https://arXiv.org/abs/0905.0973}{\path{arXiv:0905.0973}}.

\bibitem{Aceti:2014uea}
F.~Aceti, M.~Bayar, E.~Oset, A.~Martinez Torres, K.~P.~Khemchandani, J.~M.~Dias, F.~S.~Navarra and M.~Nielsen,
{Prediction of an $I=1$ $D \bar D^*$ state and relationship to the claimed $Z_c(3900)$, $Z_c(3885)$,}
Phys. Rev. D \textbf{90}, no.1, 016003 (2014).
\href{https://doi.org/10.1103/PhysRevD.90.016003}{\path{doi:10.1103/PhysRevD.90.016003}},
\href{https://arXiv.org/abs/1401.8216}{\path{arXiv:1401.8216}}.

\bibitem{Wang:2023jeu}
Z.~Y.~Wang and Z.~F.~Sun,
{Hidden strange $B_{c}$-like molecular states,}
Eur. Phys. J. C \textbf{83}, no.12, 1106 (2023).
\href{https://doi.org/10.1140/epjc/s10052-023-12283-3}{\path{doi:10.1140/epjc/s10052-023-12283-3}},
\href{https://arXiv.org/abs/2307.00803}{\path{arXiv:2307.00803}}.



\bibitem{Liu:2024pio}
W.~Y.~Liu and H.~X.~Chen,
{Hadronic Molecules with Four Charm or Beauty Quarks,}
Universe \textbf{11}, no.2, 36 (2025).
\href{https://doi.org/10.3390/universe11020036}{\path{doi:10.3390/universe11020036}},
\href{https://arXiv.org/abs/2405.14404}{\path{arXiv:2405.14404}}.

\bibitem{pdg}
R.~L.~Workman \textit{et al.} [Particle Data Group],
{Review of Particle Physics,}
PTEP \textbf{2022}, 083C01 (2022).
\href{https://doi.org/10.1093/ptep/ptac097}{\path{doi:10.1093/ptep/ptac097}}.

\bibitem{Becirevic:2013bsa}
D.~Be\v{c}irevi\'c, G.~Duplan\v{c}i\'c, B.~Klajn, B.~Meli\'c and F.~Sanfilippo,
{Lattice QCD and QCD sum rule determination of the decay constants of $\eta_c$, J/$\psi$ and $h_c$ states,}
Nucl. Phys. B \textbf{883}, 306-327 (2014).
\href{https://doi.org/10.1016/j.nuclphysb.2014.03.024}{\path{doi:10.1016/j.nuclphysb.2014.03.024}},
\href{https://arXiv.org/abs/1312.2858}{\path{arXiv:1312.2858}}.

\bibitem{Mathur:2018epb}
N.~Mathur, M.~Padmanath and S.~Mondal,
{Precise predictions of charmed-bottom hadrons from lattice QCD,}
Phys. Rev. Lett. \textbf{121}, no.20, 202002 (2018).
\href{https://doi.org/10.1103/PhysRevLett.121.202002}{\path{doi:10.1103/PhysRevLett.121.202002}},
\href{https://arXiv.org/abs/1806.04151}{\path{arXiv:1806.04151}}.

\bibitem{McNeile:2012qf}
C.~McNeile, C.~T.~H.~Davies, E.~Follana, K.~Hornbostel and G.~P.~Lepage,
{Heavy meson masses and decay constants from relativistic heavy quarks in full lattice QCD,}
Phys. Rev. D \textbf{86}, 074503 (2012).
\href{https://doi.org/10.1103/PhysRevD.86.074503}{\path{doi:10.1103/PhysRevD.86.074503}},
\href{https://arXiv.org/abs/1207.0994}{\path{arXiv:1207.0994}}.

\bibitem{Dong:2020hxe}
X.~K.~Dong, F.~K.~Guo and B.~S.~Zou,
{Explaining the Many Threshold Structures in the Heavy-Quark Hadron Spectrum,}
Phys. Rev. Lett. \textbf{126}, no.15, 152001 (2021).
\href{https://doi.org/10.1103/PhysRevLett.126.152001}{\path{doi:10.1103/PhysRevLett.126.152001}},
\href{https://arXiv.org/abs/2011.14517}{\path{arXiv:2011.14517}}.

\bibitem{Yu:2018yxl}
Q.~X.~Yu, R.~Pavao, V.~R.~Debastiani and E.~Oset,
{Description of the $\Xi _c$ and $\Xi _b$ states as molecular states,}
Eur. Phys. J. C \textbf{79}, no.2, 167 (2019).
\href{https://doi.org/10.1140/epjc/s10052-019-6665-z}{\path{doi:10.1140/epjc/s10052-019-6665-z}},
\href{https://arXiv.org/abs/1811.11738}{\path{arXiv:1811.11738}}.

\bibitem{Molina:2008jw}
R.~Molina, D.~Nicmorus and E.~Oset,
{The rho rho interaction in the hidden gauge formalism and the f(0)(1370) and f(2)(1270) resonances,}
Phys. Rev. D \textbf{78}, 114018 (2008).
\href{https://doi.org/10.1103/PhysRevD.78.114018}{\path{doi:10.1103/PhysRevD.78.114018}},
\href{https://arXiv.org/abs/0809.2233}{\path{arXiv:0809.2233}}.

\bibitem{Gulmez:2016scm}
D.~G\"ulmez, U.~G.~Mei\ss{}ner and J.~A.~Oller,
{A chiral covariant approach to $\rho\rho$ scattering,}
Eur. Phys. J. C \textbf{77}, no.7, 460 (2017).
\href{https://doi.org/10.1140/epjc/s10052-017-5018-z}{\path{doi:10.1140/epjc/s10052-017-5018-z}},
\href{https://arXiv.org/abs/1611.00168}{\path{arXiv:1611.00168}}.

\bibitem{Geng:2016pmf}
L.~S.~Geng, R.~Molina and E.~Oset,
{On the chiral covariant approach to $\rho\rho$ scattering,}
Chin. Phys. C \textbf{41}, no.12, 124101 (2017).
\href{https://doi.org/10.1088/1674-1137/41/12/124101}{\path{doi:10.1088/1674-1137/41/12/124101}},
\href{https://arXiv.org/abs/1612.07871}{\path{arXiv:1612.07871}}.

\bibitem{Du:2018gyn}
M.~L.~Du, D.~G\"ulmez, F.~K.~Guo, U.~G.~Mei\ss{}ner and Q.~Wang,
{Interactions between vector mesons and dynamically generated resonances,}
Eur. Phys. J. C \textbf{78}, no.12, 988 (2018).
\href{https://doi.org/10.1140/epjc/s10052-018-6475-8}{\path{doi:10.1140/epjc/s10052-018-6475-8}},
\href{https://arXiv.org/abs/1808.09664}{\path{arXiv:1808.09664}}.



\end{thebibliography}


\end{document}